\documentclass[conference]{IEEEtran}
\IEEEoverridecommandlockouts
\usepackage{cite}
\usepackage{amsmath,amssymb,amsfonts}
\usepackage{mathtools} 
\usepackage{tabularx,booktabs} 
\usepackage{algorithmic}
\usepackage{graphicx}
\usepackage{textcomp}
\usepackage{xcolor}
\usepackage[utf8]{inputenc}
\def\BibTeX{{\rm B\kern-.05em{\sc i\kern-.025em b}\kern-.08em
    T\kern-.1667em\lower.7ex\hbox{E}\kern-.125emX}}
\begin{document}

\title{OPF-Learn: An Open-Source Framework\\ for Creating Representative AC Optimal\\ Power Flow Datasets\\
\thanks{
This work was authored in part by the National Renewable Energy Laboratory (NREL), operated by Alliance for Sustainable Energy, LLC, for the U.S. Department of Energy (DOE) under Contract No. DE-AC36-08GO28308. The work of T. J-J was supported in part by the U.S. Department of Energy, Office of Science, Office of Workforce Development for Teachers and Scientists (WDTS) under the Science Undergraduate Laboratory Internships Program (SULI). The work of A. S. Zamzam was supported by the Laboratory Directed Research and Development (LDRD) Program at NREL. The views expressed in the article do not necessarily represent the views of the DOE or the U.S. Government. The U.S. Government retains and the publisher, by accepting the article for publication, acknowledges that the U.S. Government retains a nonexclusive, paid-up, irrevocable, worldwide license to publish or reproduce the published form of this work, or allow others to do so, for U.S. Government purposes. The work of K. Baker is supported by the National Science Foundation CAREER award 2041835.}
}

\author{\IEEEauthorblockN{Trager Joswig-Jones}
\IEEEauthorblockA{\textit{University of Washington}\\
Seattle, WA, USA  \\
joswitra@uw.edu}
\and
\IEEEauthorblockN{Kyri Baker}
\IEEEauthorblockA{\textit{University of Colorado Boulder}\\
Boulder, CO, USA \\
kyri.baker@colorado.edu}
\and
\IEEEauthorblockN{Ahmed S. Zamzam}
\IEEEauthorblockA{
\textit{National Renewable Energy Laboratory}\\
Golden, CO, USA \\
ahmed.zamzam@nrel.gov}
}

\maketitle

\begin{abstract}
Increasing levels of renewable generation motivate a growing interest in data-driven approaches for AC optimal power flow (AC OPF) to manage uncertainty; however, a lack of disciplined dataset creation and benchmarking prohibits useful comparison among approaches in the literature. To instill confidence, models must be able to reliably predict solutions across a wide range of operating conditions. This paper develops the OPF-Learn package for Julia and Python, which uses a computationally efficient approach to create representative datasets that span a wide spectrum of the AC OPF feasible region. Load profiles are uniformly sampled from a convex set that contains the AC OPF feasible set. For each infeasible point found, the convex set is reduced using infeasibility certificates, found by using properties of a relaxed formulation. The framework is shown to generate datasets that are more representative of the entire feasible space versus traditional techniques seen in the literature, improving machine learning model performance.
\end{abstract}


\section{Introduction}
A significant power systems research area addresses the challenges power system operators face when determining how to economically meet electrical demand, referred to as the optimal power flow problem. The alternating current optimal power flow (AC OPF) problem can be formulated as an optimization problem and solved for generator outputs given demand inputs, but finding the global optimum of an AC OPF model has been proven to be non-deterministic polynomial-time (NP) hard due to the nonconvexities in the problem \cite{Lavaei2012zeroduality}. Further, this problem can contain thousands of decision variables and is difficult to solve in real time. 

Additional challenges are introduced by heightened levels of variable and uncertain power generation from renewable energy resources, which necessitate faster decision making when operating grid assets. Linear approximations of network physics, typically used by grid operators to estimate power flows throughout the system, are not sufficient for ensuring overall system reliability \cite{CAPITANESCU201657ACOPFreview}, and can result in billions of dollars of economic inefficiencies annually \cite{Cain2012historyofOPF}. The “DC OPF" approximation, for instance, is used by many system operators and has been shown to result in highly inaccurate estimates of power flows and associated electricity prices \cite{Overbye2004ACDC}. There is a clear and growing need for higher-efficiency methods to solve AC OPF. 

One solution is to apply data-driven approaches. Machine learning (ML) approaches move computing complexity offline, so that near-optimal AC OPF solutions can be found in real time. This allows operators to rapidly solve the AC OPF problem and economically operate larger systems with more decision variables; however, despite promising results, there remains a lack of disciplined dataset creation and benchmarking which prohibits useful comparisons among approaches. These results rely heavily on the underlying dataset used to train a model, so it is essential that the dataset represents a wide variety of operating conditions.

Typically, AC OPF datasets used to demonstrate regression ML approaches are generated by sampling from a uniform distribution of ±10\% to 20\% around the nominal load for each load bus in a network \cite{singh2021learning,huang2021deepopfv,Fioretto2020Lagrangian,baker2019learning,Dong2020Smart-PGSIM,robson2019learning}. In \cite{Chatzos2020HighFidelityML}, a variation of this method uses two separate uniform sampling distributions to generate a lower and upper bound for the demand of each load, and then the loads are uniformly increased from the lower bound to the upper bound. Another variation of sampling around the nominal network load is shown in  \cite{Zamzam2020Extremely} by using a truncated Gaussian distribution to find a real power value for each load. In \cite{RAHMAN2021106908LearningAugmented} and \cite{rahman2020machine} a different approach is taken where sample load profiles are generated based on real historical hourly load profiles. 

These methods can result in datasets that are only representative of only a small portion of the feasible region, meaning models that are trained on these datasets might only accurately predict values for loading near the nominal loading scenario.  To gain more confidence in using these models for real systems, the data that models are trained and tested on needs to be more representative of the entire feasible space. This paper demonstrates a method for efficiently creating datasets by mapping load profiles to optimal generator set points, that are representative of the full AC OPF feasible space that will likely be encountered during network operation. The main contributions of this paper are devising a method to efficiently create datasets to facilitate the development of ML approaches to AC OPF, and showcasing the significance of using the developed datasets within a simple ML model compared with typical dataset creation methods.

\section{AC Optimal Power Flow Formulation}
A power network is defined with $N$ buses collected in set $\mathcal{N}$, with set $\mathcal{G}$ collecting the set of nodes with generators and set $\mathcal{L}$ collecting the set of load buses. The admittance matrix, denoted as $\boldsymbol{Y}$, represents the admittance of lines connecting buses in the network, with $\boldsymbol{Y}_{ij}$ indicating the (i,j) element of $\boldsymbol{Y}$. The set $\mathcal{E}$ collects direct edges representing the network lines originating from the `from' end, and the set $\mathcal{E}_t$ collects the network edges originating from the `to' end of the lines. The power injection at bus $n \in \mathcal{N}$ is denoted by $s_n = p_n + \boldsymbol{j} q_n$. Further, $s_{l,n} = p_{l,n}+ \boldsymbol{j} q_{l,n}$ and $s_{g,n} = p_{g,n}+ \boldsymbol{j} q_{g,n}$ denote the power generated at bus $n \in \mathcal{G}$ and the load demand at bus $n \in \mathcal{L}$, respectively. The voltage at bus $n$ is denoted by $v_n$. The voltage phasors, active power injections, and reactive power injections at all buses are defined as $V \coloneqq [v_n]_{n \in \mathcal{N}}$, $P \coloneqq [p_n ]_{n \in \mathcal{N}}$, and $Q \coloneqq [q_n ]_{n \in \mathcal{N}}$, respectively. Therein the AC OPF problem can be formulated as:
%
%
\begin{subequations} \label{eq:AC OPF}
\begin{align} 
\min_{V, S_g} &\quad\sum_{i \in \mathcal{G}} a_i \Re{(s_{g,i})}^2 + b_i \Re{(s_{g,i})} + c_i 
\label{eq:AC OPF_objective}\\
\text{s. t:~}&\quad (V, S_g) \in \boldsymbol{\Omega}(S_l)
\end{align}
\end{subequations}
where the nonconvex set $\boldsymbol{\Omega}(S_l)$ comprises the AC power flow equations, the generation limits, the voltage magnitude constraints, the line flow limits, and the angle limits. The AC OPF is nonconvex and, in fact, generally NP-hard~\cite{lehmann2015ac}. 

\section{Relaxation and Properties}
Because the AC OPF problem is nonconvex and hard to solve in general, many heuristic, relaxation, and linearization approaches have been proposed to solve the problem. Numerous relaxation approaches have been proposed in the literature, such as the Second-Order Cone relaxation (SOC)~\cite{Jabr2006soc}, and the Quadratic Convex (QC) relaxation~\cite{Coffrin2016QCrelaxation}. In general, a relaxed AC OPF problem can be written as:
\begin{subequations} \label{eq:SOC}
\begin{align} 
\min_{V, S_g} &\quad\sum_{i \in \mathcal{G}} a_i \Re{(s_{g,i})}^2 + b_i \Re{(s_{g,i})} + c_i 
\label{eq:SOC_objective}\\
\text{s. t:~}&\quad (V, S_g) \in \widetilde{\boldsymbol{\Omega}}(S_l)
\end{align}
\end{subequations}
where $\widetilde{\boldsymbol{\Omega}}(S_l)$ is a convex set that encompasses the nonconvex set $\boldsymbol{\Omega}(S_l)$. 
The input to the relaxed problem \eqref{eq:SOC} is the same input to the original nonconvex AC OPF problem \eqref{eq:AC OPF} which comprises the values of $p_{l,i}$ and $q_{l,i}$ at all buses $i \in \mathcal{N}$. An important property of the relaxation \eqref{eq:SOC} is that if the optimization problem \eqref{eq:SOC} is infeasible for specific values of $P_l$ and $Q_l$ , then the AC OPF problem \eqref{eq:AC OPF} is also infeasible for the same load profile $P_l$ and $Q_l$. This property provides a sufficient condition for the infeasibility of a loading situation. We will use this property to exclude loading scenarios that are provably infeasible from the input space. 

\section{Dataset Creation Methodology}
In this section, we present the main elements of the datasets generation process we used. First, we introduce the format of the data, then we present sampling mechanisms and the main steps used to increase the efficiency of the sampling process. Then we summarize the approach used to create AC OPF data that can be used for training in machine learn methods. 

\subsection{Data Format}

To support ML methods working to identify the solutions or the active constraints, for each data sample, three data containers (structures) are stored. The first contains the input information, $x \coloneqq {p_{l,n},q_{l,n}}$, ${n \in \mathcal{L}}$, which encompasses the active and reactive load demands at all buses. The second container includes the optimal solution controllable variables, $y \coloneqq {v_{g,n}^r,p_{g,n}^r}$, ${n \in \mathcal{G}}$, i.e., the voltage magnitudes at all buses in $\mathcal{G}$ and the active power injections from all the generators. Given this information, along with the inputs, one can recover the complete solution, ${V}$, by solving the AC power flow equations~\cite{Zamzam2020Extremely}. In addition, the third container includes the information required to identify active constraints. That is, we include the value of the dual variables corresponding to voltage magnitude constraints, line flow limits, and generators capacity constraints.
Note that active constraints are defined to be the constraints with nonzero dual variables.

\subsection{Initializing Input Space}
The input load space that the samples are pulled from must be initialized to a set that contains the AC OPF feasible set in the load region of interest. When working with test case networks, the minimum and maximum servable demand at each load bus must be solved for. This is done by formulating a convex AC OPF optimization problem with the objective of maximizing the load at a single bus and solving this problem for each load bus in the network. This problem can be formulated as:
\begin{subequations} \label{eq:SOC-ACOPF-Pmax}
\begin{align} 
\overline{p}_{l,i} := \arg\max_{\substack{V, S_g, S_l}}&\quad  p_{l, i} \\
\mathrm{s.t:~} & \quad (V, S_g) \in \widetilde{\boldsymbol{\Omega}}(S_l)
\end{align}
\end{subequations}

The previous optimization problem identifies an upper bound on the maximum active load, $\overline{p}_i \forall i \in \mathcal{L}$, that can be served by the generation capacity of the network. We bound the reactive power demand by choosing a lower bound on the power factor of the loads connected to the feeder. In addition, we assume that the minimum value of the loads is zero. 
On the other hand, we constrain the power factor of the load connected at each bus; hence, the reactive power injection is constrained by $0 < q_l < p_l$, which corresponds to limiting the power factor of all loads to be at least $\frac{1}{\sqrt{2}}$. We collect these constraints in a concise form by defining the input space as $A_0 x_l \leq b_0$, where:


\begin{equation}
    A_0 \coloneqq \begin{bmatrix*}[r]
    \boldsymbol{\mathrm{I}}^{|\mathcal{L}| \times |\mathcal{L}|} & [0]^{|\mathcal{L}| \times |\mathcal{L}|} \\ 
    -\boldsymbol{\mathrm{I}}^{|\mathcal{L}| \times |\mathcal{L}|} & [0]^{|\mathcal{L}| \times |\mathcal{L}|} \\
    [0]^{|\mathcal{L}| \times |\mathcal{L}|} & -\boldsymbol{\mathrm{I}}^{|\mathcal{L}| \times |\mathcal{L}|} \\
    -\boldsymbol{\mathrm{I}}^{|\mathcal{L}| \times |\mathcal{L}|} & \boldsymbol{\mathrm{I}}^{|\mathcal{L}| \times |\mathcal{L}|} \\
    [1]^{1 \times |\mathcal{L}|} & [0]^{1 \times |\mathcal{L}|}
    \end{bmatrix*},\quad b_0 \coloneqq \begin{bmatrix*}[c]
    \overline{p_l} \\ 
    [0]^{|\mathcal{L}|} \\
    [0]^{|\mathcal{L}|} \\
    [0]^{|\mathcal{L}|} \\
    \sum_{\substack{i \in \mathcal{G}}}p_{g,i} \\
    \end{bmatrix*}
\end{equation}


This linear equality describes a closed set of load profiles that will be used as the load sampling space. This unclassified space contains the relaxed feasible set and the AC OPF feasible set that could possibly be encountered during operation. 

\subsection{Sampling Input Space}
Load profiles, $x_{l}$, are sampled from the unclassified space using a method for uniformly sampling from convex polytopes. A uniform sampling method is used to ensure that the samples are representative of all AC OPF feasible load profiles that could be seen during operation. Uniform sampling from a convex polytope is a heavily studied research area \cite{kaufman1998ACHR, diaconis2012gibbs}. 
For this paper `Hit and Run' sampling, a Monte Carlo method, as described in 
\cite{CPRND}, is used to uniformly and quickly sample high dimensional polytopes. To initialize this sampler, a point within the polytope must be specified. This can be done by finding the Chebyshev center of the unclassified space. 

%
%


\subsection{Constructing Separating Hyperplanes}
Separating hyperplanes, as proposed in \cite{venzke2021efficient}, are used to classify large regions of the input space as infeasible. These hyperplanes are based on infeasibility certificates that are created when a load profile is sampled that is infeasible for the  relaxed AC OPF. When an infeasible load profile, $\Hat{x}_l$, is sampled, the nearest input, $x^*_l$, that is feasible for the convex relaxed AC OPF is found by solving \eqref{eq:SOC-FNFP}.
%
\begin{subequations} \label{eq:SOC-FNFP}
\begin{align} 
x^*_{l} := \arg\min_{\substack{V, S_g, {x}_l}}\quad & \| {x}_{l} - \hat{x}_l \|_2 \\
\mathrm{s.t:~}\quad & (V, S_g) \in \widetilde{\boldsymbol{\Omega}}({x}_l)
\end{align}
\end{subequations}
Once $x^*_l$ is found, an infeasibility certificate can be constructed if $x^*_l \ne \hat{x}_l$. The vector from the infeasible sample to the nearest feasible point, $\overrightarrow{n} \coloneqq x_l^* - \Hat{x}_l$, defines the normal vector of the new hyperplane. This normal vector and the nearest feasible point then define the hyperplane as $A_l x_l \leq b_l$, where $ A_l \coloneqq \overrightarrow{n}$ and $b_l \coloneqq \overrightarrow{n} x_l^*$. $A_l$ and $b_l$ are added as new rows to $A$ and $b$, respectively, to include the hyperplane in the definition of the load space polytope to reduce the volume of the input space.

\subsection{Summary of Approach}
\begin{figure}[htbp]
\centerline{\includegraphics[width=0.5\textwidth]{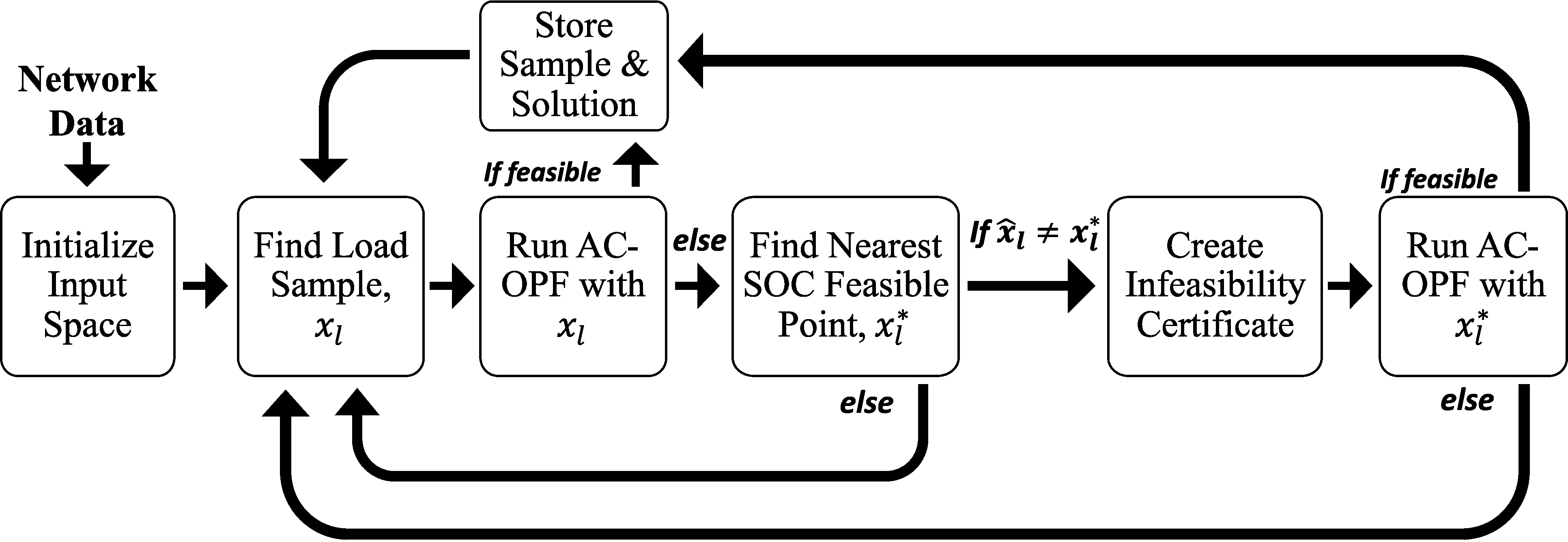}}
\caption{OPF-Learn procedure flowchart.}
\label{flowchart}
\end{figure}

\begin{figure}[htbp]
\centering
\includegraphics[width=0.5\textwidth]{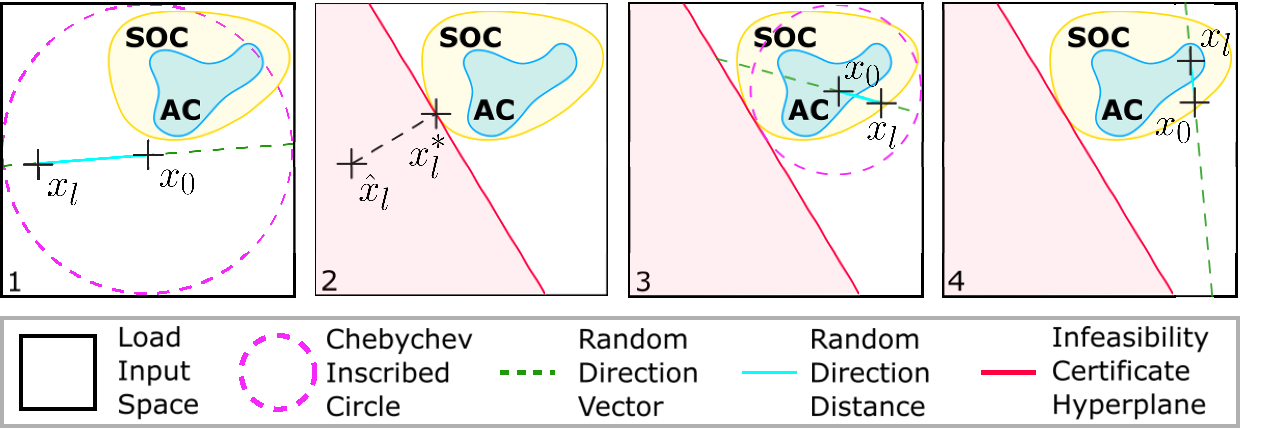}
\caption{(1) Find the Chebyshev center to use as the initial point, $x_0$. Generate a random direction vector and travel a random distance along this vector to find a new load sample, $x_l$. (2) Check if $x_l$ is AC OPF feasible. If it is not feasible, find the nearest relaxed feasible point, $x_l^*$. Because $\hat{x}_l \ne {x}_l^*$ define a new infeasibility certificate at $x_l^*$ with normal, $\vec{n} = \hat{x}_l - {x}_l^*$. (3) Gather a new sample, $x_l$, as in Step 1. Check if the new sampled load is AC OPF feasible. Here, it is not, so the nearest relaxed feasible point is found. $\hat{x}_l = x^*_l$ so discard this sample. (4) Sample a new load profile, $x_l$, as in Step 1, but starting from the last point, now $x_0$. Check if $x_l$ is AC OPF feasible. $x_l$ is AC OPF feasible, so store $x_l$ and its AC OPF optimal solution.}
\label{fig:example}
\end{figure}

This methodology finds load samples by uniformly sampling from a convex set, the input space, which contains the AC OPF feasible set. Samples are then tested for AC OPF feasibility and are added to the dataset if they are feasible. The convex set is reduced throughout sampling by constructing separating hyperplanes to increase the likelihood of sampling feasible load profiles. Fig.~\ref{flowchart} shows the process used to create the AC OPF datasets. Fig.~\ref{fig:example} shows an example of the sampling processes resulting in the construction of an infeasibility certificate.
\vspace{-4.5mm}
\section{Simulation and Results}
Here, we discuss features of the OPF-Learn datasets, and we compare these datasets to ones created using a typical dataset creation method as seen in the literature. Using both datasets, we train and test neural networks to see the possible implications of using a typical dataset creation method when evaluating a ML approach to solve the AC OPF problem.

\subsection{Dataset Creation}
We evaluate our proposed AC OPF dataset creation method on 5 PGLib-OPF networks
\cite{pglib} up to 118 buses. For each of these 5 networks, we create datasets with $N=10,000$ samples. These datasets are created with our Julia package, OPFLearn.jl.
PowerModels.jl \cite{powermodels} with IPOPT \cite{ipopt} 
is used to find locally optimal feasible solutions to the nonconvex AC OPF problem formulated in JuMP \cite{JuMP}. QC-relaxed AC OPF problems for finding maximum feasible bus demands and nearest feasible load profiles were formulated in JuMP and solved with IPOPT. Initially, the stronger semi-definite relaxation \cite{BAI2008383sdp} was used, but the run times were found to be too long with existing conic solvers to justify its use. The QC relaxation was selected over the SOC relaxation because it is a tighter relaxation, as demonstrated in \cite{Coffrin2016QCrelaxation}. Note that for case118, the maximum loads used to initialize the sampling input space were set to be twice the nominal load instead of the found maximum loads to decrease the dataset creation run time. In practice, it was found that many infeasibility certificates of the OPF-Learn sampling method are required to reduce the input space to one with a good portion of feasible solutions with larger networks, especially when the initial feasible space contains feasible solutions with an unrealistically large amount of active power demand at a single load bus, as seen with networks that have more than 100 buses.

A typical dataset for each network was also created with $N=10,000$ samples. These datasets were created by sampling from a uniform distribution $\mathcal{U}(0.8 x_{0,i}, 1.2 x_{0,i})$ for each $i \in \mathcal{L}$, where $x_{0} = (P_0, P_0)$ is the nominal load at bus i. For each load sample generated that was found to be AC OPF feasible, the sample and AC OPF solution were stored. 

Comparing the datasets created from these two methods shows that the typical datasets have significantly fewer unique active sets than the OPF-Learn datasets. Table~\ref{table:uas} shows the number of unique active sets in each dataset.
\newcolumntype{Y}{>{\centering\arraybackslash}X}
\newcolumntype{R}{>{\raggedleft\arraybackslash}X}
\begin{table}
\centering
\caption{Total number of unique active sets found by OPF-Learn and the typical method in 10,000 feasible samples.}
\vspace{-1.5mm}
\begin{tabularx}{0.8\columnwidth}{p{0.12\columnwidth} *{2}{Y}}
\toprule
 & OPF-Learn Dataset & Typical Dataset \\
\midrule
\multicolumn{1}{r}{case5}   & 774  & 13 \\
\multicolumn{1}{r}{case14}  & 6938 & 7 \\
\multicolumn{1}{r}{case30}  & 9931 & 7 \\
\multicolumn{1}{r}{case57}  & 9958 & 33 \\
\multicolumn{1}{r}{case118} & 9980 & 2731 \\
\bottomrule
\end{tabularx}\label{table:uas}
\vspace{-10pt}
\end{table}

\begin{figure}[htbp]
\centerline{\includegraphics[width=0.4\textwidth]{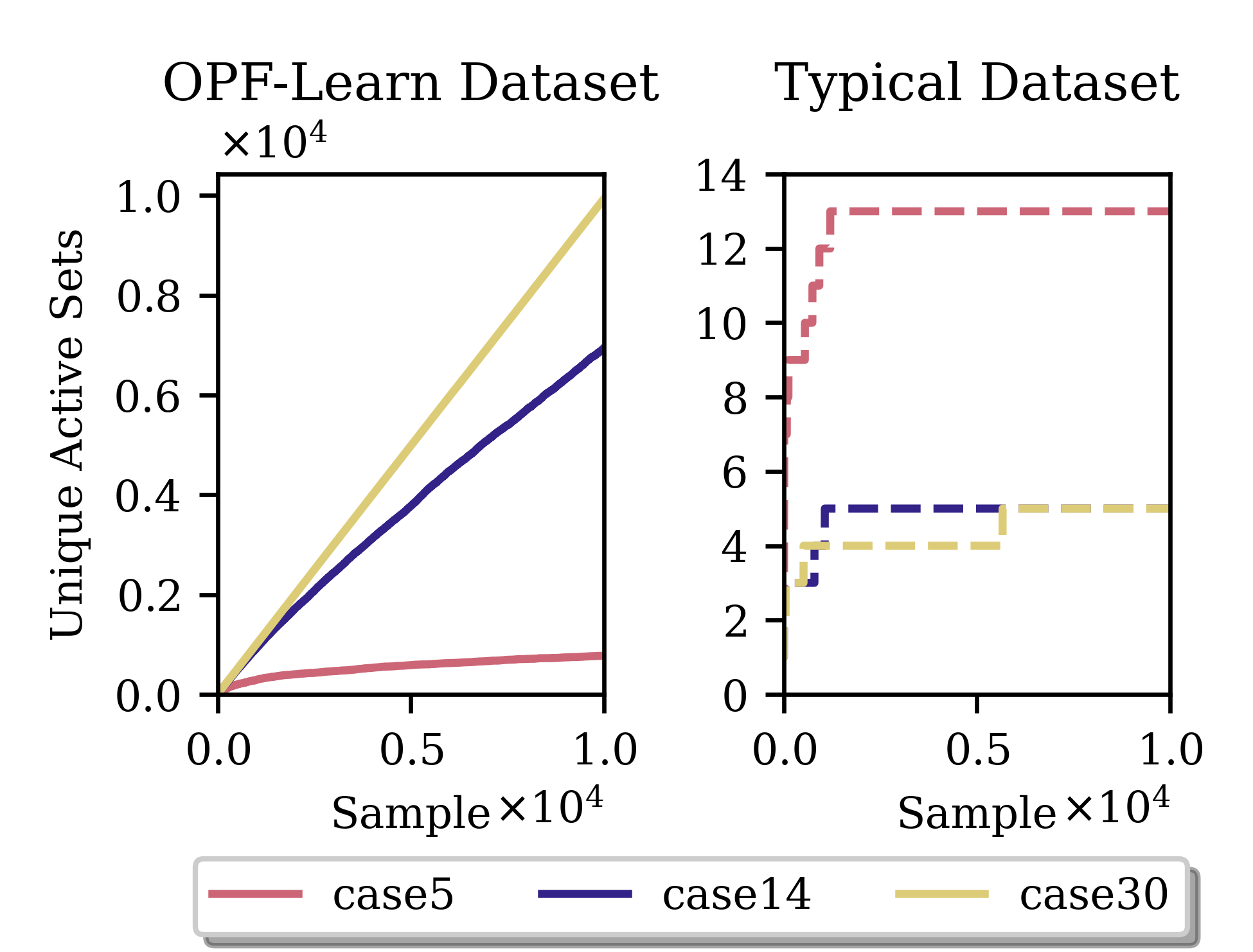}}
\vspace{-3pt}\caption{The number of feasible unique active sets found throughout dataset creation with OPF-Learn and with the typical method. Note the difference in y-axis scales due to the large disparities in the unique active sets found.}
\label{unique_active_sets}
\vspace{-10pt}
\end{figure}

Fig.~\ref{unique_active_sets} shows the rate at which unique active sets were found for each feasible sample found. The number of unique active sets in the OPF-Learn datasets were not found to converge in these 10,000 sample datasets for cases larger than case5. This indicates that there are likely more unique active sets in the AC OPF feasible space, and a larger dataset would be required to classify the entire feasible space within the input space. For the larger systems, neither the OPF-Learn sampling method or the typical sampling method are seen to converge for these 10,000 sample datasets, but fewer new unique active sets are found with the typical sampling method.

The larger number of unique active sets found using OPF-Learn indicates that the typical dataset is only representative of a smaller portion of the AC OPF feasible space around the nominal load, while the OPF-Learn dataset contains points representative of more loading conditions. Because the mappings of loads to AC OPF solutions within a unique active set are much simpler, the mapping being learned by an ML model trained on a typical dataset can only be extrapolated accurately within these unique active sets.

\subsection{Training Neural Networks}
We used these datasets to train and test two sets of neural network (NN) models to demonstrate how ML models trained on a simpler typical datasets perform on datasets that are more representative of the entire AC OPF feasible space that could be seen during operation. For both NN models, we used a NN with three hidden layers with a sigmoid activation functions between all hidden layers. The width of the first two hidden layers is equal to the number of inputs, $2|\mathcal{L}|$, and the width of the last hidden layer is equal to the number of outputs, $|\mathcal{G}|$. The input data are in per unit and are additionally normalized before being input to the NN model.

For each test network, four NN models are trained. The models are trained on a training set of either OPF-Learn data or typical data, and they predict either $P_g$ or $|V_g|$ for each generator in the network. Separate models for each output are used to simplify evaluating which output variable is associated with the most significant portion of the error. The OPF-Learn dataset and the typical dataset are split into train and test datasets with an 80-20\% train-test split. For both an output of $P_g$ and an output of $|V_g|$, one model is trained on the OPF-Learn training dataset, and the other is trained on the typical training dataset. The NN models were implemented using the Python-based TensorFlow software library \cite{tensorflow2015-whitepaper} and trained using the Adam optimizer with a mean squared error cost function. Once trained, we tested these models on test sets created using both methods to see how well they were able to predict the AC OPF solution for a given load profile.

\begin{table}
\caption{$P_g$ and $|V_g|$ NN model test mean squared error results for NN model trained on an OPF-Learn dataset or a typical dataset for all PGLib test networks.}
\vspace{-1.5mm}
\begin{tabularx}{\columnwidth}{p{0.12\columnwidth} *{4}{Y}}
& & $P_g$ models \\
\toprule
 & \multicolumn{2}{c}{
 \begin{minipage}[c]{0.365\columnwidth}%
 OPF-Learn Training Dataset
 \end{minipage}
 }  
 & \multicolumn{2}{c}{
 \begin{minipage}[c]{0.365\columnwidth}%
 Typical Training Dataset
 \end{minipage}
 }  \\
\cmidrule(lr){2-3} \cmidrule(l){4-5}
 \quad Test Network & 
 OPF-Learn Test Dataset & Typical \quad Test Dataset & 
 OPF-Learn Test Dataset & Typical \quad Test Dataset \\
\midrule
\multicolumn{1}{r}{case5}   & 2.17E-2 & 1.86E-3 & 1.33E+0 & 9.08E-6 \\
\multicolumn{1}{r}{case14}  & 2.75E-4 & 1.01E-4 & 3.94E-2 & 9.41E-7 \\
\multicolumn{1}{r}{case30}  & 1.55E-4 & 5.46E-4 & 8.17E-3 & 1.60E-8 \\
\multicolumn{1}{r}{case57} & 2.37E-1 & 1.73E-1 & 7.54E-1 & 2.39E-2 \\
\multicolumn{1}{r}{case118} & 6.97E-2 & 2.35E-1 & 4.47E-1 & 4.47E-3 \\
\bottomrule \\

& & $|V_g|$ models \\
\toprule
 & \multicolumn{2}{c}{
 \begin{minipage}[c]{0.365\columnwidth}%
 OPF-Learn Training Dataset
 \end{minipage}
 }  
 & \multicolumn{2}{c}{
 \begin{minipage}[c]{0.365\columnwidth}%
 Typical Training Dataset
 \end{minipage}
 }  \\
\cmidrule(lr){2-3} \cmidrule(l){4-5}
 \quad Test Network & 
 OPF-Learn Test Dataset & Typical \quad Test Dataset & 
 OPF-Learn Test Dataset & Typical \quad Test Dataset \\
\midrule
\multicolumn{1}{r}{case5}   & 3.43E-5 & 5.53E-6 & 9.51E-3 & 2.49E-8 \\
\multicolumn{1}{r}{case14}  & 1.29E-5 & 1.60E-5 & 5.72E-4 & 2.58E-7 \\
\multicolumn{1}{r}{case30}  & 1.26E-5 & 2.22E-5 & 2.39E-4 & 1.08E-8 \\
\multicolumn{1}{r}{case57} & 2.96E-4 & 2.51E-4 & 3.25E-4 & 1.13E-5 \\
\multicolumn{1}{r}{case118} & 3.80E-4 & 1.05E-2 & 4.61E-4 & 7.63E-5 \\
\bottomrule
\end{tabularx}\label{table:NNresults}
\vspace{-10pt}
\end{table}

The test results for the different NNs are shown in Table ~\ref{table:NNresults}. For all cases, the typical training set NNs saw a significantly larger error when tested on the OPF-Learn test set compared to the typical test set for both $P_g$ and $|V_g|$. These results show that the typical dataset-trained $P_g$ NNs performs worse on the OPF-Learn test dataset than on the typical test dataset by orders of magnitude from $10^1$ to $10^5$. The largest increase in error occurred with case30, with the error increasing by 5.09E+5 times. In comparison, the OPF-Learn-trained NNs had an increase only up to 2.77E+1 times in error, as shown with the $|V_g|$ case118 NN. These large increases in error with the typical-dataset trained NNs indicates the possibility that ML models demonstrated in the literature trained on similar typical datasets might not see the same level of performance reported in papers if they are tested on a dataset that is more representative of the AC OPF feasible space. 

With the $P_g$ case30 typical training set NN, the maximum sample error, measured as the sum of the absolute value of the difference of predicted and actual AC OPF solutions, was found to increase from 4.64E-03 to 8.38E-01, showing that some load samples in the OPF-Learn dataset produce significantly more suboptimal solutions than would be predicted with the typical test set. Similar increases in the maximum solution suboptimality were seen for all typical dataset-trained models, and while increases in the maximum error were seen with the OPF-Learn-dataset trained models, these increases in suboptimality were smaller.

%
%
%
\section{Conclusion}
This paper presented OPF-Learn, an open-source software tool for efficient AC OPF dataset creation to facilitate the development of ML methods to solve the AC OPF problem. The sampling approach starts by considering a space that includes all load profiles.
Then, properties of relaxation approaches are used to shrink the space to efficiently create feasible samples. The demonstrations on PGLib networks show that the generated datasets are more diverse and representative than typical sampling approaches, which only vary loads around base values. In addition, assessing the performance of ML models trained on typical datasets results in ML models that have limited generalization. The open-source package delivers a timely requirement for the fair benchmarking of ML methods developed to tackle the AC OPF problem.

%
%
%
%
%
%

\bibliographystyle{IEEEtran}
\bibliography{refs} 

\vspace{12pt}

\end{document}